\documentclass[twoside,twocolumn]{article}
\usepackage[super,sort&compress,comma]{natbib} 
\usepackage{times,mathptmx}
\usepackage{sectsty}
\usepackage{balance} 

\usepackage{graphicx} 
\usepackage{lastpage}
\usepackage[format=plain,singlelinecheck=false,font=small,labelfont=bf,labelsep=space]{caption} 
\usepackage{fancyhdr}
\usepackage{amsmath}
\newcommand{\ket}[1]{|#1\rangle}
\newcommand{\ad}{a^\dagger}
\newcommand{\bra}[1]{\langle#1|}

\newcommand{\x}{\mathbf{x}}

\newcommand{\kLSH}{{\sc $k$-Local-Spin Hamiltonian}}
\newcommand{\tLSH}{{\sc $2$-Local-Spin Hamiltonian}}

\newcommand{\REP}{{\sc $N$-Rep}}
\newcommand{\UF}{{\sc Universal Functional}}
\newcommand{\ES}{{\sc Electronic Structure}}
\newcommand{\HF}{{\sc Hartree-Fock}}

\newcommand{\qma}{{QMA}}

\newcommand{\p}{{P}}

\newcommand{\id}{\mathbf{1}}
\newcommand{\X}{\sigma^x}
\newcommand{\Y}{\sigma^y}

\newcommand{\Z}{\sigma^z}

\renewcommand{\eqref}[1]{equation~(\ref{#1})}
\begin{document}

\thispagestyle{plain}
\fancypagestyle{plain}{
\fancyhead{}
\renewcommand{\headrulewidth}{1pt}}
\renewcommand{\thefootnote}{\fnsymbol{footnote}}
\renewcommand\footnoterule{\vspace*{1pt}%
\hrule width 3.4in height 0.4pt \vspace*{5pt}} 
\setcounter{secnumdepth}{5}

\makeatletter 
\def\subsubsection{\@startsection{subsubsection}{3}{10pt}{-1.25ex plus -1ex minus -.1ex}{0ex plus 0ex}{\normalsize\bf}} 
\def\paragraph{\@startsection{paragraph}{4}{10pt}{-1.25ex plus -1ex minus -.1ex}{0ex plus 0ex}{\normalsize\textit}} 
\renewcommand\@biblabel[1]{#1}            
\renewcommand\@makefntext[1]%
{\noindent\makebox[0pt][r]{\@thefnmark\,}#1}
\makeatother 
\renewcommand{\figurename}{\small{Fig.}~}
\sectionfont{\large}
\subsectionfont{\normalsize} 

\fancyfoot{}
\fancyfoot[RO]{\footnotesize{\sffamily{1--\pageref*{LastPage} ~\textbar  \hspace{2pt}\thepage}}}
\fancyfoot[LE]{\footnotesize{\sffamily{\thepage~\textbar 1--\pageref*{LastPage}}}}
\fancyhead[LE]{J. D. Whitfield et al.}
\fancyhead[RE]{}
\fancyhead[RO]{Computational Complexity in Electronic Structure}
\fancyhead[LO]{}
\renewcommand{\headrulewidth}{1pt} 
\renewcommand{\footrulewidth}{1pt}
\setlength{\arrayrulewidth}{1pt}
\setlength{\columnsep}{6.5mm}
\setlength\bibsep{1pt}

\twocolumn[
  \begin{@twocolumnfalse}
	  \noindent\LARGE{\textbf{Computational Complexity in Electronic Structure}}
\vspace{0.6cm}

\noindent\large{\textbf{James Daniel Whitfield,\textit{$^{a,b,c}$}, Peter John Love\textit{$^{d}$} and Al\'an Aspuru-Guzik\textit{$^{e}$}}}\vspace{0.5cm}

\vspace{0.6cm}

\noindent 
\normalsize{In quantum chemistry,
the price paid by all known efficient model chemistries is either the
truncation of the Hilbert space or uncontrolled approximations.  Theoretical 
computer science suggests that these restrictions are not mere shortcomings of the 
algorithm designers and programmers but could stem from the inherent difficulty 
of simulating quantum systems. Extensions of computer science and information 
processing exploiting quantum mechanics has led to new 
ways of understanding the ultimate limitations of computational power.  
Interestingly, this perspective helps us understand 
widely used model chemistries in a new light.  In this article, the 
fundamentals of computational complexity will be reviewed and motivated from the vantage point of 
chemistry. Then recent results from the computational complexity 
literature regarding common model chemistries including Hartree-Fock and 
density functional theory are discussed.
}
\vspace{0.5cm}
 \end{@twocolumnfalse}
  ]

\section{Introduction}

\footnotetext{\textit{$^{a}$~Vienna Center for Quantum Science and Technology (VCQ),  Boltzmanngasse 5, 1090 Vienna, Austria. E-mail: James.Whitfield@univie.ac.at}} 
\footnotetext{\textit{$^{b}$~Columbia University, Physics Department, 538 West 120th Street, New York, NY, USA.}}
\footnotetext{\textit{$^{c}$~NEC Laboratories America, 1 Independence Way, Princeton, NJ, USA.}} 
\footnotetext{\textit{$^{d}$~Haverford College.  370 Lancaster Avenue, Haverford, PA, USA. Email: plove@haverford.edu }}
\footnotetext{\textit{$^{e}$~Harvard University, Department of Chemistry and Chemical Biology, Cambridge, MA, USA. Email: aspuru@chem.harvard.edu }}

Quantum chemistry is often concerned with solving the Schr\"odinger equation for 
chemically-relevant systems such as atoms, molecules, or nanoparticles.  
By solving a differential and/or eigenvalue 
equation, the properties of the system and the dynamics of the state are obtained.  Examples of properties include: equilibrium geometries, the dissociation energy of molecules, and the vibrational frequencies.

The difficulty stems from the accuracy required and the apparent exponential growth of the computational cost with both the number of electrons and the quality of the description of the system. For practical applications, the accuracy required is typically orders of magnitude smaller than the total energy of the system. As a concrete example, the Carbon atom has total electronic energy of about 37.8 Hartrees while the energy of a Carbon-Hydrogen bond is only 0.16 Hartrees.  Solving the full eigenvalue equation takes on the order of $n^3$ operations for an $n\times n$ matrix. However, when describing interacting many-electron systems, the dimension of the matrix increases exponentially with the number of electrons.

Consequently, the computational methods of electronic structure in chemistry are aimed at circumventing exact diagonalization, in the context of electron structure, called the full configuration method.  Avoiding exact diagonalization has led to a wide range of computational methods~\cite{Szabo96,Helgaker00,Koch01,Cramer04} for computing properties of chemical interest.
These methods have recently~\cite{Aspuru05,Brown10,Kassal11,Whitfield11,Yung12} begun to include \emph{quantum simulation} following Feynman's suggestion\cite{Feynman82} to use quantum computers as simulators. Subsequent development of these ideas in quantum chemistry has led to new proposed methods utilizing quantum computational techniques\cite{Wu02,Kassal08,Veis10,Whitfield11,Biamonte11} and proof-of-principle experiments~\cite{Lu12,Aspuru12}.   However, questions about when and where one would expect a quantum computer to be useful\cite{Aaronson08,Love12} have not been fully answered. Ideally, computational complexity can provide some answers about when, where, and why quantum computers would be useful for chemistry. At the same time, it could also help formalize intuitive  understanding of when we can expect reliable results from traditional computational methods.

Many results in computational complexity have interchanged classical and quantum computers fluidly leading to new results on the complexity of computing properties of quantum systems~\cite{Osborne12}.    We review some recent results appearing in the computational complexity literature that touch on why electronic structure calculations are difficult.
Our hope is to encourage future investigations into quantitative understanding of difficult instances of electronic structure calculations.

A similar, but much shorter, discussion of computational complexity in quantum chemistry by~\citet{Rassolov08} appeared in 2008 and provided many conjectures that have since been proven or extended. 

This perspective assumes exposure to second quantization and mixed states in quantum mechanics.  Standard bra-ket notation is used when referring to quantum states. All the necessary concepts from computational complexity and quantum computation are briefly introduced and motivated to make the article as self-contained as possible.  A key omission from the present review is a number of computational complexity results ~\cite{Schuch07,Schuch08} concerning quantum-information based wave function ansatzes such matrix product states, density matrix renormalization group and their generalizations~\cite{Verstraete08}.
These methods are becoming accepted into the mainstream of computational chemistry~\cite{Chan08,Chan11,Chan12}, but do not yet have the widespread availability of the selected methods included in the present work.

\section{Worst case computational complexity for chemists}\label{sec:intro}

The purpose of this section is to provide a quick but precise introduction to computational complexity concepts\cite{Sipser97,Arora09}. The aim is to set the stage for the results subsequently reviewed. 

Computational complexity is the study of how resources required to solve a problem change with its size.   For instance, \emph{space complexity} is the scaling of memory requirements with the problem size, but, in this article, we focus on \emph{time complexity} which investigates how the running time of the computation changes as the problem size increases. 

Computational chemists often informally discuss common-place concepts in computer science such as the complexity classes of polynomial-time problems (P) and non-deterministic polynomial-time problems (NP). For instance, it is sometimes stated that Hartree-Fock has a runtime which scales as the third power of the number of basis functions.  This is scaling disregards difficult instances of the calculation where Hartree-Fock does not converge. Such instances require manual intervention to tweak the algorithm used or adjust the convergence thresholds in a case-by-case fashion. 

In computer science, often the focus is on \emph{worst-case complexity} where the  most difficult instances of a problem are used to classify the problem's complexity. This is has been one of the major areas of theoretical computer science, although work on average case complexity does exist~\cite{BenDavid92,Bogdanov06}. In this article, the complexity of the problems discussed are characterized by the worst-case scaling.

\subsection{Time complexity in equivalent computer models}
A proper measure of the \emph{time complexity} of an algorithm is how many basic operations (or how much time) it takes to solve problems of increasing size. Conventionally, a computational problem is described as easy or tractable if there exists an efficient algorithm for solving it, i.e.~one that scales polynomially, $O(n^k)$ with fixed $k$ and input size $n$.\footnote{The notation, $f(x)=O(g(x))$, implies that function $f(x)$ is dominated by $g(x)$ for asymptotically large values of $x$.  To indicated that $f(x)$ is asymptotically larger than $g(x)$, we write $f(x)=\Omega(g(x))$.  If $f(x)$ is dominated by and dominates function $g(x)$, then we write $f(x)=\Theta(g(x))$.} Otherwise, the problem is considered intractable. This is an asymptotic definition that may not capture the full utility of the algorithm.  For example, an asymptotically efficient algorithm may run slower than an asymptotically inefficient for small or fixed size problem instances.
Nevertheless, this asymptotic classification of algorithms has proved useful. From a theoretical computer science perspective, the division allows for considerable progress to be made without considering the minutiae of the specific system,  implementation, or domain of application.

From a practical standpoint, Moore's\cite{Moore65} law states the density of transistors in classical computers doubles every two years. Thus, for a fixed computational time, if the computer cannot solve an instance when using a polynomial-time algorithm, one need not wait long as the exponential growth of computing power will reasonably quickly  overtake the polynomially large cost.  However, if the algorithm runs in exponential-time, one may be forced to wait several lifetimes in order for an instance to become soluble even if Moore's law continues indefinitely.  Complicating matters, the exponential growth according to Moore's law is expected to cease sometime this century; hence the recent emphasis on 
quantum computation.

The time complexity can be characterized using any computationally equivalent model.  In the context of computer science, \emph{equivalent} means that one model can simulate the other with an overhead that scales polynomially as a function of system size.  These mappings respects the boundary between efficient and inefficient algorithms described earlier.  
Turing machines and circuits are two typical models discussed in the context of computational complexity.

\begin{figure}[t]
	\begin{center}
		\includegraphics[width=\columnwidth]{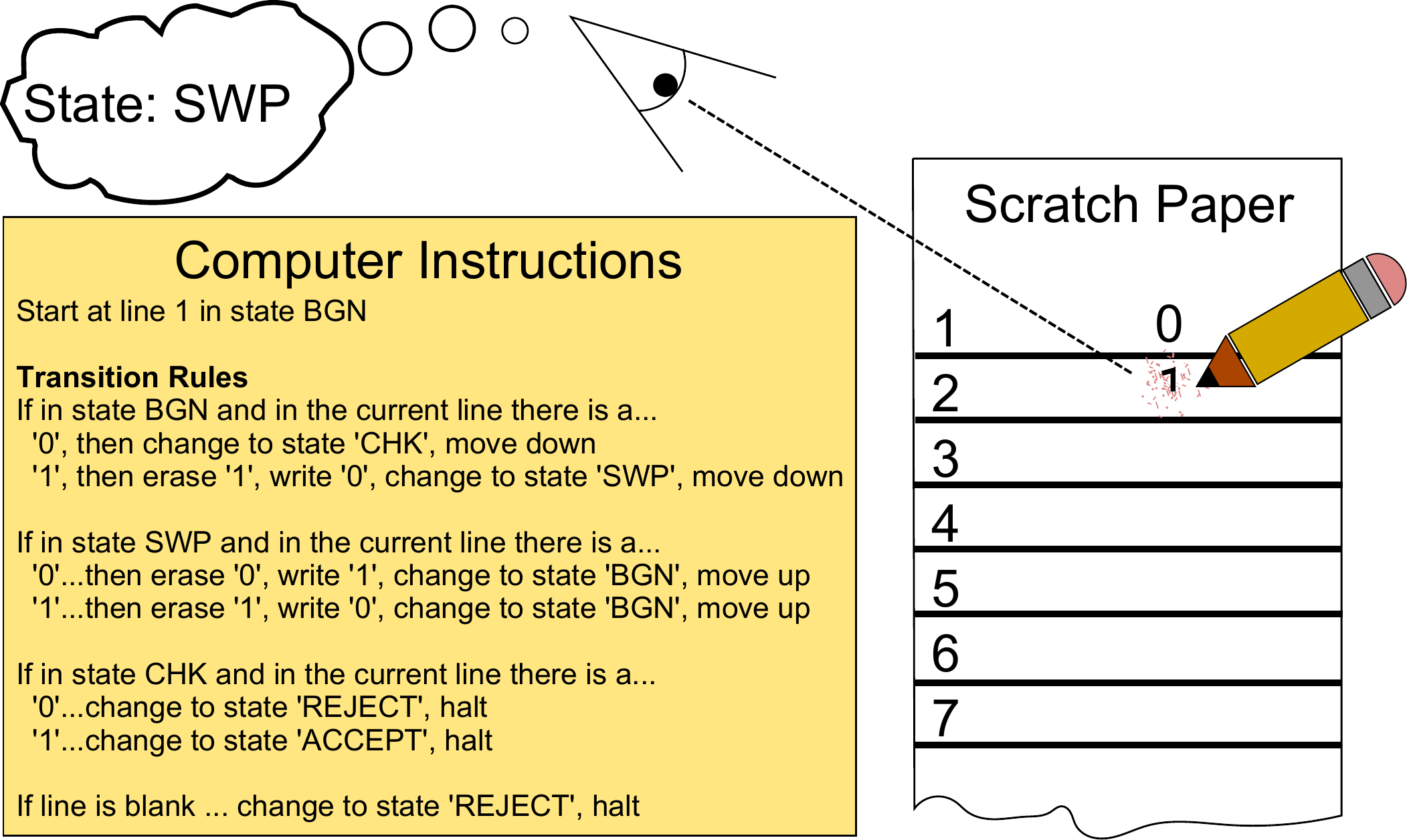}
	\end{center}
	\caption{Illustration of a Turing machine that accepts input strings beginning with `01' or `10.' The computer is specified by the set of states (modes of operation), an alphabet of symbols for the scratch space, and the transition rules.  The computer performing an algorithm reads the symbol on the current line, then based on the computer's current state and the given transition rules, the computer changes the symbol of the current line, changes its current state and moves either up or down. The computer halts when it enters the ACCEPT or REJECT states which indicate the output of the computation.  In the figure, the possible states are \{BGN, CHK, SWP, REJECT, ACCEPT\}, the alphabet is \{`0',`1'\}, and the transition rules are listed in the box to the left. Depicted is the second step of a computation that began with input `10'.}
\label{fig:turing}
\end{figure}

The Turing computer or \emph{Turing machine} was introduced by Alan~\citet{Turing36} in 1936 before transistors and electronic circuits and formalizes the idea of a computer as a person to whom computational instruction could be given. \citet{Turing36} introduced the concept in order to answer David Hilbert's challenge to decide if a polynomial has roots which are integers using a ``finite number of operations.'' Turing proved that this was not possible using his Turing machine. An illustration depicting the salient features of the Turing machine is given in Fig.~\ref{fig:turing}.

The circuit model of computation has been more widely used when generalizing to the quantum setting although the quantum Turing machine formulation does exist~\cite{Deutsch85,Bernstein93}. The time complexity in the circuit model is characterized the number of circuit elements (or gates) used to compute the accept/reject boolean value corresponding to an input.  So long as a universal set of gates, e.g.~NAND, FAN-OUT, and  FAN-IN for classical circuits, are used, the time complexity will only differ by a polynomial factor.  The key caveat to consider when using this model is the notion of uniformity.  To properly define an algorithm in the circuit model, a way of generating circuits for all input sizes must be given concisely.  A circuit family for an algorithm is \emph{uniform} if there is a polynomial-time algorithm that specifies the circuit given the input size.  

\subsection{Computational problems and problem instances}

In computer science, computational problem often refers to decision problems which are collections of yes/no questions that an algorithm can decide.  Each question is called a \emph{problem instance} and the answer is called the \emph{solution} to that instance. Although we focus on decision problem, in many situations the decision problem can be used to extract numerical values by asking sufficiently many yes/no questions.

A collection of decision questions with affirmative answers such as, ``Is 149 prime?'' or ``Is 79 prime?,'' is called a \emph{language}. Languages define decision problems where only the accept instances are included in the language. If an algorithm \emph{accepts} (returns ``True'' or outputs $1$) on all strings contained in a language then the device is said to \emph{recognize} the language.  If the computer is also required to halt on all inputs, then it is said to \emph{decide} the language. The difference being, a Turing machine may recognize a language but fail to halt on some instances, i.e.~run forever. The problem of deciding if a Turing machine algorithm halts is called the \textsc{Halting} problem.  This problem has a rich and storied history beginning with the first paper of \citet{Turing36}.  In Fig.~\ref{fig:turing}, the algorithm decides the language of strings beginning with `01' or `10.'

As an example from chemistry, the computational problem of deciding if a molecule has a dipole moment, {\sc Dipole}, is a collection of questions: ``Does $x$ have a dipole moment?'' Here, $x$ is a string representing the molecule in each instance. The questions `Does BCl$_3$ have a dipole moment?', and `Does NH$_3$ have a dipole moment?' are instances of this computational problem. The string $x$=``BCl$_3$'' is not in the language {\sc Dipole} since ``BCl$_3$'' does not have dipole moment while ``NH$_3$'' is in the language since it has a dipole moment.  A reasonable modification of the problem would be `Does $x$ have a dipole moment greater than 0.1 Debyes?' so that small dipole moments may be ignored. 

It may also be necessary to promise that $x$ represents a molecule in a specific way such that ill formatted inputs can be ignored.  This is accomplished using \emph{promise problems} where the promise would be that $x$ is indeed a string that properly encodes a molecule. Promise problems also arise in order to account for issues of numerical or experimental precision as illustrated in the next paragraph.  

Promise problems also play a key role throughout the remainder of this text since we are discussing physical properties where infinite precision is neither required nor expected.  As an illustration, we use an example from thermochemistry, where language $A$ is the set of strings corresponding to (ideal) gases with heat capacity at constant pressure, $C_p$, less than some critical value. Imagine you have an unreasonable lab instructor, who gives you a substance that has heat capacity extremely close to the critical value that decides membership in $A$.  It may take a large number of repetitions of the experimental protocol to be confident that the substance belongs or does not belong to language $A$. A reasonable lab instructor would announce at the beginning of lab that all the substances he is handing out for the experiment are promised to be at least one Joule per Kelvin away from the critical value.  Given this promise, the student will be able to complete their lab in a single lab period.  Without such a promise, it may take the student several lab periods to repeat the experiment in order to establish a sufficiently precise value of the heat capacity to decide the instance. Instead when using a promise, if the student has a compound that violates the promise then the instructor would give full credit for either answer.  

More formally, in the usual decision problems, the problem is defined using one language, $L$, specifying the accept instances.  The language of reject instances is all strings not in $L$.  Promise problems differ in that they are specified using two languages: one for the accept instances, $L_{accept}$, and a separate language for the reject instances, $L_{reject}$. If an instance is not in either of these languages, then the promise is violated and the computation can terminate with any result.

Reusing the \textsc{Dipole} example, if we can reformulate the problem to find out about non-zero dipole moments as a promise problem to account for experimental error.  Now the computational task of \textsc{Dipole} is: `Given a molecule, $x$, a trial value for the dipole $d_T$, and an error tolerance $\delta$,  decide if the dipole moment of $x$ is greater than $d_T+\delta$ or less than $d_{T}-\delta$, promised that the dipole moment is not between $d_T\pm\delta$.'  In this case, there are two languages defining the problem $L_{d>d_T+\delta}$ and $L_{d<d_T-\delta}$.  If a molecule has a dipole moment of exactly $d_T$ it would violate the promise and the experimenter or computer does not need to answer or can respond with any answer.  The value $\delta$  allows us to meaningfully define problems in the presence of errors resulting from imperfect experimental measurements or from the finite precision of a computing device.

\subsection{Computation reductions}
The idea of computational reduction is at the heart of classifying computational
problems. Reducibility is a way of formalizing the relationship between problems.  Essentially, it asks, ``If I can solve one problem, what other problems can I solve using the same resources?''  

There are two main types of reductions: Turing reductions and Karp reductions.  For the reduction to be useful, they must be limited to polynomial time. The polynomial-time \emph{Turing reduction}, also called the Cook reduction, of problem $A$ to problem $B$ uses solutions to multiple instances of $B$ to decide the solutions for instances of $A$.  The solutions to instances of $B$ are provided by an \emph{oracle} and each questions is called a \emph{query}.  Algorithms deciding $A$ which require only polynomial queries of the oracle for $B$ are efficient whenever the oracle for $B$ answers in polynomial time. 

The other type of reduction is called a \emph{Karp reduction} or polynomial transformation.  If, instead of an oracle, there is an embedding of instances of problem $A$ into instances of problem $B$, then an efficient solution for all instances of problem $B$ imply efficient solutions for instances of problem $A$.  This transformation is a Karp reduction.  When $A$ can be \emph{reduced} to $B$, under either Karp of Turing reductions, it is denoted $A\leq B$.

To illustrate the difference, we use examples from thermodynamics.  Consider trying to determine if the heat capacity at constant pressure, $C_p$, of an ideal substance is less than some critical value (language $A$).  Now, suppose an oracle, probably in the form of a bomb calorimeter, has been provided for deciding if the heat capacity at constant volume, $C_V$, of a given substance is above or below a critical value.  Call language $B$ the set of strings labeling substances with $C_V$ below this value.  By adding the number of moles times the ideal gas constant to the critical value, one can determine membership in language $A$ via the formula $C_p=C_V+nR$. Since each instance of $A$ can be embedded into an instance of $B$, language $A$ is Karp reducible to $B$ and we can write $A\leq B$.

Suppose instead, an oracle for evaluating if, at fixed pressure, the internal energy, $U$, of the given substance at a given temperature is less than some critical value (language $C$) is given.  Then by evaluating the internal energy at two different temperatures, the heat capacity at constant pressure can be bounded by numerically estimating the derivative.  Because the reduction has to use the oracle for $C$ more than once to decide instances of $A$, language $A$ is Turing reducible to $C$ and $A\leq C$.

\subsection{Basic complexity classes}
Equipped with the basic definitions from computer science, we now introduce the concept of computational complexity classes. This will give some insights into why quantum chemistry is difficult. We will introduce six basic complexity classes to classify the time complexity of decision problems.  The first three complexity classes are characterized by algorithms that can decide instances in time proportional to a polynomial of the input size; the other classes are characterized by polynomial time verification of problem instances.

 In table~\ref{tbl:polytime}, the three polynomial time complexity classes are listed. First, {P}, is the class of all decision problems with instances that can be accepted or rejected by a Turing machine in polynomial time.  If the Turing machine has access to an unbiased random number generator, then the complexity class of problems that can be decided in polynomial time is called BPP. The term  ``bounded error'' refers to the requirement that the probability of acceptance and of rejection must be bounded away from half so that repetition can be employed to boost the confidence in the answer. An important class of problems falling into this complexity class are efficient Monte Carlo simulations often grouped under the umbrella term quantum Monte Carlo~\cite{Hammond94,Foulkes01,Lester09} used for computing electronic structure in chemistry. 
Lastly, the complexity class BQP is characterized by problems soluble in polynomial time with quantum resources.  In the quantum computational model~\cite{Nielsen00}, the quantum algorithm is conceptually simpler to think of as a unitary circuit, $U$, composed of unitary circuit element that affect, at most, two quantum bits. As mentioned earlier, the number of gates used determines the time complexity.  The outcome of the algorithm with input $\ket{Input}$ is ``accept'' with probability $|\bra{Accept}U\ket{Input}|^2$. Similarly, for the ``reject'' cases.
\begin{table}[t]
\small
\caption{ Polynomial time complexity classes are separated by the resources the computer has access to while the non-deterministic polynomial time complexity classes are characterize by the resources of polynomial time computational verifiers}
  \label{tbl:polytime}
  \begin{tabular*}{0.5\textwidth}{@{\extracolsep{\fill}}ccl}
    \hline
    Class & Name & \begin{tabular}{c}Computer type\\ taking only poly.~time\end{tabular} \\
    \hline
    P   & \begin{tabular}{c}Polynomial Time\\\end{tabular} & \begin{tabular}{l} Turing machine\end{tabular} \\[3ex]
    BPP &\begin{tabular}{c} Bounded Error\\Probabilistic Polynomial\\Time\end{tabular} & \begin{tabular}{l}Turing machine with access\\to random number generator\end{tabular} \\[6ex]
    BQP & \begin{tabular}{c} Bounded Error\\Quantum Polynomial\\Time\end{tabular}  & \begin{tabular}{l}Turing machine with access\\to quantum resources\end{tabular}\\
    \hline
    \hline
    Class & Name &\begin{tabular}{c} Verifier's computer type\\taking only poly.~time\end{tabular} \\
    \hline
    NP   & \begin{tabular}{c}Non-deterministic\\Polynomial Time\end{tabular} & \begin{tabular}{l}Turing machine\end{tabular} \\[3ex]
    MA &\begin{tabular}{c} Merlin-Arthur\end{tabular} & \begin{tabular}{l}Turing machine with access\\ to random number generator\end{tabular} \\[3ex]
    QMA & \begin{tabular}{c} Quantum\\Merlin-Arthur \end{tabular}  & \begin{tabular}{l}Turing machine with access\\ to quantum resources\end{tabular}\\
    \hline
  \label{tbl:nondettime}
  \end{tabular*}
\end{table}

So far, the discussion has centered on complexity classes containing problems that are solvable in polynomial time given access to various resources: a computer (P), a random number generator (BPP), and a controllable quantum system (BQP).  Now, our attention turns to the class of problems that can be computed non-deterministically.  The original notion of non-determinism is a Turing machine whose transition rule maps the computer state and the tape symbol to any number of possible outputs.  In essence, this is a computer that can clone itself at will to pursue all options at once.  
NP is the class of problems that could be solved in polynomial time by such a computer.  Whether a deterministic polynomial time computer can be used to simulate such a computer is a restatement of the famous open question in computer science: ``Does {P} = {NP}?'' This question was selected as a millennium problem by the Clay Mathematics Institute which has offered a one million dollar prize for a correct proof of the answer. 

\begin{figure*}
\begin{center}
\includegraphics{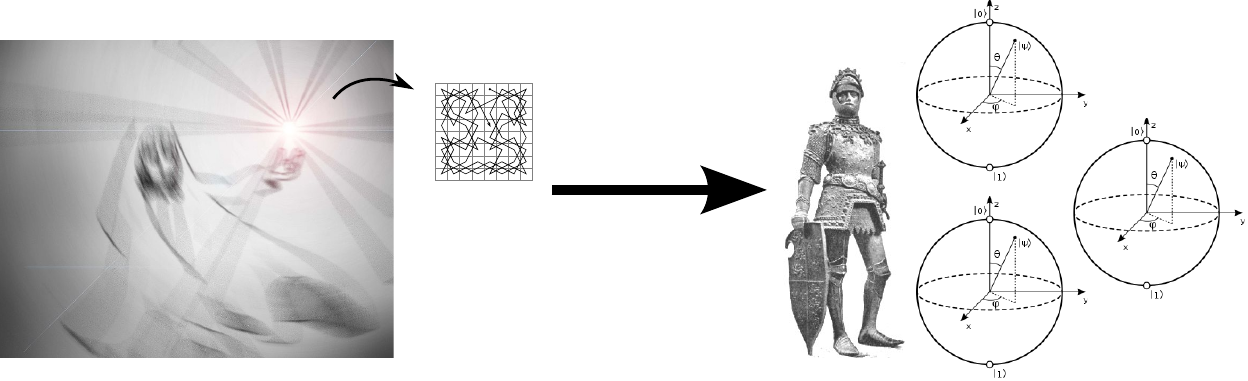}
\end{center}
\caption{An example of the non-deterministic complexity class QMA.  Non-deterministic problems can be thought of as games.  In the Merlin-Arthur games, the proof verifying the validity of input $x$ is magically (hence non-deterministically) given by a wizard Merlin who is prone to deception.  The verifier, called Arthur, is trying to decide if he should accept or reject input $x$. When Merlin gives Arthur a valid proof that verifies $x$ should be accepted, Arthur should accept with high probability (\emph{completeness}) and when Merlin gives him an proof of something incorrect, he should be able to spot the inaccuracy (\emph{soundness}). In the drawing since Arthur has access to quantum bits (qubits), the complexity class depicted is  QMA. Figure composed from images found on Wikipedia under the Creative Commons License. }
\label{fig:MA}
\end{figure*}

Rather than resorting to imaginary computers, the non-deterministic classes can be defined by considering verification of solutions which have been obtained in some non-deterministic way, see Fig.~\ref{fig:MA}.   For every ``accept'' instance, $x$, there exists at least one proof state $y$ such that the verifier  returns ``accept'' in polynomial time. If $x$ is should be rejected, then for every proof state $y$, the verifier should output ``reject'', again, in a polynomial amount of time. Each of the non-deterministic classes listed in table~\ref{tbl:nondettime}, are characterized by the computational power of the verifier.   Note that {P} is a subset of {NP} because any problem that can be easily solved can be easily verified.

\subsection{Completeness and hardness}
The goal of classifying computational problems into complexity classes motivates introduction of the terms \emph{hard} and \emph{complete}. The classification of a computational problems as hard for a complexity class means if an algorithm can solve this problem, then all problems in the class can be solved efficiently.  In other words, this problem contains all problems in the class either via Karp or Turing reductions. More precisely,
a language $B$ is \emph{hard} for a complexity class {CC} if every language, $A$, in {CC} is polynomial-time reducible to $B$, i.e.~$A\leq B$.

Complete problems for a class are, in some sense, the hardest problems in that class. These problems are both a member of the class and are also hard for the class.  That is,
a language $B$ is \emph{complete} for a complexity class {CC} if $B$ is in the complexity class CC and for every other language $A$ in {CC}, $A\leq B$.

The simplest illustration of the difference between hard and complete 
computational problems is the distinction between optimization problems and decision problems.
Optimization problems such as finding saddle points or minima in 
energy landscapes are frequently encountered in computational chemistry,
however these problems are not in the complexity class NP.  However, if one can perform optimization quickly, then responding with the answer to yes/no questions about the solution would be simple. Thus, optimization problems can be classified as NP-hard but not NP-complete since the computational task of optimization is, in a sense, harder than just answering yes or no.

\section{Electronic structure and other Hamiltonian problems}\label{sec:cchem}
Now, armed with the key ideas from computer science, we return to our original inquiry into why quantum chemistry is hard.  We answer this questions by exploring the computational complexity of three different widely used methods for computing electronic energy in computational chemistry: Hartree-Fock, two-electron reduced density matrix methods, and density functional theory.

Before delving into the specific complexities of these problems, we first establish notation by reviewing necessary concepts from quantum chemistry, then discussing computational problems concerning classical and quantum spin Hamiltonians. We end the section with mappings between systems of electrons and spin systems.

\subsection{Quantum chemistry and second quantization}
In quantum chemistry, the annihilation, $\{a_k\}$, and creation operators, $\{a_k^{\dagger}\}$, correspond respectively to removing and adding an electron into one of $M$ single particle wave functions, $\{\phi_k(\mathbf{x_1})\}_{k=1}^M$. The single-particle wave functions are called \emph{orbitals} and the set of orbitals is typically called the \emph{basis set}.  To include the electron spin, the single particle orbitals $\phi_k$ are functions of spatial coordinates and a spin variable which are collectively denoted $\mathbf{x}$.

The electronic spin will play an important role when discussing the connections between systems of electrons and systems of quantum spins.  Since electrons are spin-$\frac12$ particles, the electron spin is either up or down. Accordingly, there are $M/2$ orbitals with spin up and $M/2$ with spin down. Unless explicitly noted, the summation over orbitals includes a summation over the spatial and spin indices of the orbitals.

Anti-symmetry of the $N$-electron is enforced by the canonical anti-commutation relations for fermions, i.e.~electrons,
\begin{align}
&[a_p, \;a_q]_+= a_pa_q+a_qa_p=0,& &[a_p, \;\ad_q]_+=\delta_{pq}\id.
\label{:CCR}
\end{align}
In chemistry, typically the electronic structure is the primary concern, but if interested in vibrational structure, we would have to consider bosonic canonical commutation relations: $[b_p,b_q]=b_pb_q-b_qb_p=0$ and $[b_p,b_q^\dag]=\delta_{pq}\id$. 
The vacuum state, $\ket{vac}$, is the normalized state that the creation/annihilation operators act on and it represents the system with no particles. 
  
 Acting on the vacuum state with a strings of $N$ distinct creation operators yields $N$-electron wave functions.  The most general $N$-electron state within a basis set is, 
 \begin{equation}
	 \ket{\Psi}=\sum_K^M C_K\; \ad_{K_1}\ad_{K_2}\cdots\ad_{K_N}\ket{vac},
	 \label{eq:Psi}
 \end{equation}
with $K=(K_1,\cdots,K_N)$ and the complex valued $C_K$ are constrained such that $\sum_K |C_K|^2=1$. To convert between creation and annihilation operators and the coordinate representation, one uses field operators: $\hat{\phi}(\mathbf{x})=\sum_k\phi_k(\mathbf{x})a_k$ as in 
$\Psi(\mathbf{x_1},\cdots,\mathbf{x_N})=\bra{vac}\hat{\phi}(\mathbf{x_1})\cdots\hat{\phi}(\mathbf{x_N})\ket{\Psi}/\sqrt{N!}$. The factor of $N!$ accounts for permutation symmetries in the summation. Each state $\Psi$ or $\ket{\Psi}$ is called an $N$-electron \emph{pure state}.
A valid $N$-electron \emph{mixed state},
\begin{equation}
	\rho^{(N)}=\sum p_i \ket{\Psi_i}\bra{\Psi_i}
\end{equation}
has $\sum p_i=1 $ where each $\ket{\Psi_i}$ is an $N$-electron wave function. 
The $k$-th order reduced density matrix, abbreviated $k$-RDM, is defined using the field operators: 
\begin{align}
&\rho^{(k)}(\x_1,\x_1',\cdots,\x_k,\x_k')\nonumber\\
=&\frac{1}{(N)_k}\bra{\Psi}\hat{\phi}^\dag(\x_k')\cdots\hat{\phi}^\dag(\x_1')\hat{\phi}(\x_1)\cdots\hat{\phi}(\x_k)\ket{\Psi}\label{eq:purePT}\\
=&\frac{1}{(N)_k}\textrm{Tr}\left[\hat{\phi}(\x_1)\cdots\hat{\phi}(\x_k)\;\rho^{(N)
}\;\hat{\phi}^\dag(\x_k')\cdots\hat{\phi}^\dag(\x_1')\right]\label{eq:mixedPT}
\end{align}
With $(N)_k=N!/(N-k)!$, the reduced density matrices are normalized to unity and, in \eqref{eq:mixedPT}, the trace sums over the expectation values of states from a complete set of $N-k$-electron wave functions.

Lastly, we define the primary computational chemistry problem encountered in computation of electronic structure.  

\paragraph*{Computational Problem: }~\ES.  The inputs are the number of electrons, $N$, a set of $M$ orbitals, a static configuration of nuclei, a trial energy, $E_T$, an error tolerance $\delta$ and Hamiltonian,
\begin{equation}
	H_{elec}=T_e+V_{ee}+V_{eN}=\sum_{ij}(T_{ij}^e+V^{eN}_{ij})\ad_ia_j+V_{ijkl}^{ee}\ad_{i}\ad_ja_ka_l
	\label{eq:elec}
\end{equation}
with $T_e$ the electronic kinetic energy operator, $V_{ee}$ the electron-electron interaction operator and $V_{eN}$ the electron-nuclear interaction. The task is to decide if the ground state energy is less than $E_T-\delta$ or greater than $E_T+\delta$ promised that the energy is not between $E_T\pm\delta$.

\paragraph*{}
Methods related to getting approximate solutions to \ES~have complexities ranging from NP-complete to QMA-complete as shown in sections~\ref{sec:HF} and~\ref{sec:reduced}. The complexity of \ES~itself is not yet clear.  In the literature, only Hamiltonians with more flexibility, namely a local magnetic field, have been demonstrated to be QMA-complete as shown see section~\ref{sec:DFTQMA}.

\subsection{Classical and quantum spin Hamiltonians}

The problem of estimating the ground-state energy of Hamiltonians of different forms lies at the intersection of computer science, physics and chemistry.  In this section, we define computational problems related to both classical and quantum spin Hamiltonians.   Spin systems have long been known to provide fertile ground for complexity theory~\cite{Barahona82,Istrail00,Osborne12,Kitaev02}.  
Given a Hamiltonian, deciding if a there exists a spin configuration, 
e.g. ~$(\uparrow\uparrow\downarrow\cdots)$, that satisfies a certain property, e.g. has energy below a certain threshold, can be difficult even when checking the property for each configuration is easy.
Computations designed to decide properties of this sort are often in the complexity class NP.  If, additionally, that particular property can embed instances of any other NP properties, then that property corresponds to an NP-complete computational problem. 

\paragraph*{Computational Problem:}~\textsc{Ising}. The inputs are $N$ classical spins with possible values $\pm1$, trial energy, $E_T$, and Ising Hamiltonian 
\begin{equation}
	H_{ising}=-\sum_{\langle ij\rangle}^{neighbors}J_{ij}S_iS_j
\end{equation}
having $J_{ij}$ either $0$ or $\pm1$. The task is to decide if there is a configuration of spins $c=(s_1,s_2,\cdots,s_N)$ such that the energy of the Ising Hamiltonian is less that $E_T$ or if all configurations have energy above $E_T$. 

\vskip1ex
Unlike the other Hamiltonian problems discussed in this review, \textsc{Ising}  does not require a promise on the precision because the energy values of the Hamiltonian are integer-valued instead of continuous.

Although some instances of the Ising lattice can be solved analytically, e.g. one-dimensional Ising lattices or the solution by \citet{Onsager44} for two-dimensional Ising lattices with uniform couplings, \citet{Barahona82} showed that problem {\sc Ising} is NP-hard when interactions are restricted to nearest neighbors on a $2\times L\times L$ lattice.  Since this model is only of tangential interest to electronic structure, details of this proof are omitted.  Since the ``accept'' instances can be verified in polynomial time given a configuration $c$ with energy less than $E_T$, this problem is in the complexity class NP and hence, is NP-complete. Further results show that all non-planar lattices, even in two-dimensions, are NP-complete~\cite{Istrail00}.

To provide examples of Hamiltonian problems that are complete for the quantum analogue of NP, we turn to the quantum analogues of the Ising Hamiltonian: quantum spin Hamiltonians.  To define similar quantum Hamiltonians with QMA-complete properties, it turns out that one only needs spins with angular momentum of $\frac12$ which are often called \emph{qubits}.  The interactions are now expressed using tensor products of the Pauli sigma matrices with matrix representations
\begin{align}
	&\X=\left[ \begin{array}{rr}
		0&1\\1&0
	\end{array}\right],& &\Y=\left[ \begin{array}{rr}
		0&-i\\i&0
	\end{array}\right],& &\Z=\left[ \begin{array}{rr}
		1&0\\0&-1
	\end{array}\right].
\label{eq:pauli}
\end{align}
Additionally, $\sigma^0$ is the identity matrix.

A widely studied class of qubit Hamiltonian problems whose computational complexity was first investigated by~ \citet{Kitaev02} is the problem of computing, to polynomial accuracy, the ground state energy of $k$-spin local Hamiltonians. This set of problems is referred to as $k$-\textsc{Local Hamiltonian} in the computational complexity literature following~\cite{Kitaev02}, but we use the name \kLSH~to emphasize the nature of the Hamiltonian.
\paragraph*{Computational Problem:}~\kLSH. Given $k$-spin-local Hamiltonian acting on $N$ spins,
\begin{equation}
	H_{kQMA}=-\sum^m_{\substack{D=(d_1,\ldots,d_k)\\C=(c_1,\ldots,c_k)}}J_{C,D}\;\sigma_{c_1}^{d_1}\otimes\sigma_{c_2}^{d_2}\otimes\cdots\sigma_{c_k}^{d_k}
	\label{eq:kQMA}
\end{equation}
where $d_i\in\{x,y,z,0\}$, $c_i$ labels a spin, there are $m=$poly($N$) terms, and $|J_{C,D}|\leq 1$, decide if the ground state energy is less that $E_0-\delta$ or if the ground state energy is greater than $E_0+\delta$ with $\delta<1/\textrm{poly}(N)$ promised that is is not between $E_0\pm\delta$. %
\vskip1ex
Note that the problem is defined so that the error scales relative to the number of non-zero terms $m$ in \eqref{eq:kQMA}.  To understand this, consider the following set of instances with error tolerance $\delta=0.5$. Consider Hamiltonian, $H$, with ground state energy 17.7 and Hamiltonian, $H'$, with energy 17.6. As it stands, the two Hamiltonians cannot be distinguished.  However, by including each term in $H$ and $H'$ ten additional times, the ground state energies are now 177 and 176, which can be resolved at error tolerance $\delta$.  By considering $m$ as a polynomial in $N$, one cannot indefinitely rescale the energy to effectively shrink the error tolerance.

The first demonstration of a QMA-complete problem\cite{Kitaev02} required $k=5$. Subsequently $k$ was reduced\cite{Kempe06} to $2$. Finally, the problem was shown~\cite{Oliveira08} to remain QMA-complete even when $J_{C,D}$ is non-zero only if two spins are spatially adjacent on a 2D lattice. The problem can be further restricted by including only a limited set of two-spin interactions and still remain QMA-complete~\cite{Biamonte08}. Complexity results concerning qubit Hamiltonians as well as other variants with higher dimensional spins and more restrictive lattices were recently reviewed by~\citet{Osborne12}.

\subsection{Relationships between spin systems and electronic systems}\label{sec:relate}
Since chemists are not necessarily interested in qubit or quantum spin systems, we now discuss connections to systems indistinguishable electrons.  In this subsection, we will examine how to embed a system of spins into an electronic Hamiltonian and how to embed electronic systems into spin Hamiltonians. Lastly, we use these connections to give an example of a fermionic system (although not \ES) that is QMA-complete. 

Given an electronic Hamiltonian, there is a \emph{orbital pair pseudo-spin} mapping~\cite{Cleveland76,Auerbach94} that is used to embed qubit models to the ground state of the electronic systems.
To embed $N$ spins, there must be $M=2N$ orbitals and $N$ electrons, i.e.~half-filling, and each quantum spin is identified with a pair of fermionic modes.
The embedding requires translating each spin $\ket{q_i}=\alpha\ket{\uparrow_i}+\beta\ket{\downarrow_i}$ to fermionic operators: $\alpha\; \ad_{i\uparrow}+\beta\;\ad_{i\downarrow}$.  The Pauli operators appearing in the spin Hamiltonian, then become single fermion terms, e.g.~$h_{ij}a_i^\dag a_j$. As important examples consider, 
\begin{subequations}\label{eq:fermionicZ}
\begin{align}
\X_i&= \ket{\downarrow_i}\bra{\uparrow_i}+\ket{\uparrow_i}\bra{\downarrow_i}&
&\leftrightarrow
      &&\ad_{i\downarrow}a_{i\uparrow}+      \ad_{i\uparrow}a_{i\downarrow}\\
\Y_i&= i(\ket{\downarrow_i}\bra{\uparrow_i}-\ket{\uparrow_i}\bra{\downarrow_i})&
&\leftrightarrow
       &&i(\ad_{i\downarrow}  a_{i\uparrow}-\ad_{i\uparrow}a_{i\downarrow})\\
\Z_i&= \ket{\uparrow_i}\bra{\uparrow_i}-\ket{\downarrow_i}\bra{\downarrow_i}&
&\leftrightarrow
	   &&\ad_{i\uparrow}a_{i\uparrow}-\ad_{i\downarrow}a_{i\downarrow}
\end{align} 
\end{subequations}
 Taking tensor products of the Pauli matrices leads to two-fermion terms, e.g. $h_{ijkl}a_i^\dag a_j^\dag a_k a_l$.  The final concern is preventing double occupancy within a pair of sites which would invalidate the pseudo-spin interpretation. This is handled~\cite{Auerbach94,Liu07,Schuch09} by including an additional term which penalizes invalid electronic configurations. For each pair of  electronic modes, the two-fermion penalty 
\begin{equation}
 P_i=C\;a_{i\uparrow}^{\dagger}a_{i\uparrow}a_{i\downarrow}^{\dagger}a_{i\downarrow}
 \label{eq:penalty}
\end{equation}
is added to the electronic Hamiltonian. Because penalty $P=\sum_i P_i$ commutes with the fermionic Pauli matrices, the ground state still corresponds to the solution of the spin Hamiltonian. The constant $C$ can be selected as a low order polynomial in the system size to ensure that the ground state remains in a valid pseudo-spin state\footnote{More precisely, the norm of the Hamiltonian $\|H\|$ is upper bounded by the sum of the individual terms.  Since there are, at most, $O(N)^4$ terms the norm of the total Hamiltonian must scale less than a fourth order polynomial in the system size.}. Note that we have not explicitly relied on the anti-commutation properties and a nearly identical construction exists for bosonic systems~\cite{Wei10}.

A second connection is given using well established techniques developed to translate certain spin systems to simpler non-interacting fermionic systems that can be exactly solved\cite{Jordan28,Lieb61,Dutta10}. 
The Jordan-Wigner transform\cite{Jordan28} provides this connection by mapping fermions to spin operators such that the canonical anti-commutation relations are preserved.   The Jordan-Wigner transform between $N$ fermionic creation and annihilation operators and the Pauli matrices acting on $N$ spins is given by
\begin{subequations}\label{:JW}
\begin{eqnarray}
	a_j\Leftrightarrow{\id}^{\otimes j-1}\otimes \sigma^{+} \otimes {\Z}^{\otimes N-j}
\label{subeq:JW(dest)}\\
    a_j^\dagger\Leftrightarrow{\id}^{\otimes j-1}\otimes \sigma^{-} \otimes {\Z}^{\otimes N-j}
\label{subeq:JW(crea)}
\end{eqnarray}
\end{subequations}
where $\sigma^+= \frac{\X+i\Y}{2}=\ket{\downarrow}\bra{\uparrow}$ and $\sigma^-=\frac{\X-i\Y}{2}=\ket{\uparrow}\bra{\downarrow}$. The qubit state $\ket{\uparrow\dots\uparrow}$ corresponds to the vacuum state and the string of $\Z$ operators, preserve the commutation relations in \eqref{:CCR} since $\Z$ and $\sigma^\pm$ anti-commute.  More sophisticated generalizations of the Jordan-Wigner transform, reduce the number and the spatial extent of the spin-spin interactions~\cite{Bravyi02,Verstraete05,Batista01}.

The orbital pair pseudo-spin mapping and the Jordan-Wigner transformation complement each other and will be used repeatedly throughout the remainder of the article.  The orbital pair pseudo-spin mapping requires carefully engineered penalties and fixes the total number of orbitals.  Thus, it is primarily useful when translating spin systems to electronic systems.  By contrast, the Jordan-Wigner transformation is primarily useful in the other direction, that is, when translating an arbitrary electronic system to a quantum spin system.  The next subsection gives a concrete example of how these connections are employed when studying the computational complexity of electronic systems. 

\subsection{Generic local fermionic problems are QMA-complete}\label{sec:generic}
The orbital pair construction allows one to immediately ascertain that the ground state energy decision problem for Hamiltonians containing two-fermion interactions, 
\begin{equation}
	H_{2f}=\sum h_{ij}a^\dag_ia_j+h_{ijkl}a_i^\dag a_j^\dag a_ka_l,
	\label{eq:2f}
\end{equation} 
is QMA-hard~\cite{Liu07}.  This follows via a Karp reduction using the orbital pair pseudo-spins as in~\eqref{eq:fermionicZ}. Since only two-spin interactions in \eqref{eq:kQMA} are required for QMA-completeness~\cite{Kempe06}, each spin-spin interaction term translates to two-fermion terms under the pseudo-spin mapping. Since the pseudo-spin construction can be extended to bosonic systems, the two-boson ground state energy decision problem is also QMA-hard~\cite{Wei10}.

The ground state energy of the electronic state can be verified to be above or below $E_T\pm\delta$ by a BQP quantum computer given the proof state. This implies that the problem is QMA-complete. We sketch the idea relying on well known results about BQP quantum simulation of chemical systems~\cite{Kassal11,Brown10,Yung12}. The Jordan-Wigner transformation is used to translate the two-electron Hamiltonian into a qubit Hamiltonian.  If Merlin provides the fermionic ground state\footnote{The verifier requires multiple copies of the state to ensure that the state has exactly $N$ electrons.}, the energy of corresponding qubit state can be determined through simulation\cite{Ortiz01,Somma02,Whitfield11} of the evolution under the qubit version of the two-electron Hamiltonian.  This quantum evolution can be simulated efficiently on a quantum computer~\cite{Lloyd96,Berry07}. The resulting evolution in the time-domain is Fourier transformed to extract the energy of the ground state~\cite{Kitaev95,Wu02,Aspuru05,Whitfield11}, allowing the verifier to determine whether to accept or reject the instance.  This argument carries through for the bosonic case as well using the bosonic equivalent of the Jordan-Wigner transform~\cite{Wei10}.

\section{Hartree-Fock}\label{sec:HF}
Hartree-Fock is one of the most important quantum chemistry techniques as it typically recovers about 99\% of the total electronic energy.  Hartree-Fock is known to be a weak approximation in many instances, but it is the basis for more sophisticated (post-Hartree-Fock) methods which improve upon the Hartree-Fock wave function.  Furthermore, Hartree-Fock provides the mathematical framework for the widely adopted notion of molecular orbitals used throughout chemistry. 

The implementation of the Hartree-Fock algorithm requires evaluating and manipulating $O(M^4)$ two-electron integrals.  When this computation dominates the runtime, the algorithm scales as somewhere between $O(M^2)$ and $O(M^3)$. However, since Hartree-Fock solves a nonlinear eigenvalue equation through an iterative method \cite{Szabo96}, the convergence of self consistent implementations is the key obstacle that prevents the worst case scaling from being polynomial.  The computational complexity result proving that the worst case scaling cannot be polynomial unless \p=NP was provided in an unpublished appendix of \citet{Schuch09} available on the arXiv preprint server.

For the purposes of this article, the Hartree-Fock procedure can be succinctly explained as the minimization of the energy of an $N$-electron system given $M$ basis functions with the restriction that in the expansion found in \eqref{eq:Psi}, all $C_K$ are zero except one.  Explicitly,
\begin{equation}
	E_{HF}=\min_{\substack{\Psi=\prod_i(b^\dag_i)^{k_i}\ket{vac},\\ \sum k_i=N}}\langle \Psi|H_{elec} |\Psi\rangle
	\label{eq:HF1}
\end{equation}
Here, the single Fock state corresponding to the minimal value of $E_{HF}$ is called the 
Hartree-Fock state: $\Psi_{HF}$. The optimized set of creation operators,
\begin{equation}
	b_i^\dagger=\sum_j C_{ij}\ad_j
	\label{eq:MO}
\end{equation}
place and remove electrons from the set of \emph{molecular orbitals}, $\psi_j(\x)=\sum_j^M C_{ij}\phi_i(\x)$.  The formal computational problem \HF~will be defined analogous to the other Hamiltonian problems with a promise given to account for precision.

\paragraph*{Computational Problem:}~\textit{\HF.} The inputs are the number of electrons, $N$, a set of $M$ orbitals, a trial energy, $E_T$, an error tolerance, $\delta<1/\textrm{poly}(N)$, and a two-electron Hamiltonian, c.f.~\eqref{eq:2f}. The sum of the absolute values of the coefficients is required to scale less than a polynomial in $N$.  The task is to decide if the Hartree-Fock energy $E_{HF}=\bra{\Psi_{HF}}H\ket{\Psi_{HF}}$ is less than $E_T-\delta$ or greater than $E_T+\delta$ promised that $E_{HF}$ is not between $E_T\pm\delta$.

\subsection{\HF~is NP-complete}\label{thm:HF} 

To prove that any other NP problem can be mapped to the HF problem (NP-hardness), we can use a Karp reduction to embed instances of \textsc{Ising} into instances of HF~\cite{Schuch09}.  Using the fermionic version of $\sigma^z$ given in \eqref{eq:fermionicZ}, $H_{ising}$ only requires two-electron interactions,
\begin{eqnarray}
	&&S_iS_j= \Z_i \Z_j\nonumber\\&=& (\ad_{i\uparrow}a_{i\uparrow}-\ad_{i\downarrow}a_{i\downarrow})(\ad_{j\uparrow}a_{j\uparrow}-\ad_{j\downarrow}a_{j\downarrow})\label{eq:ZZ}
\end{eqnarray}
The satisfying assignment of the \textsc{Ising} instance is some spin configuration $\ket{s_1s_2\cdots s_N}$ for $N=2L^2$ with $s_i$ as either $\uparrow$ or $\downarrow$. The correct and exact Hartree-Fock solution for the $N$-electron wave function should assign $b^\dagger_i=\ad_{i\uparrow}$ when $s_i=\uparrow$ and $b^\dagger_i=\ad_{i\downarrow}$ when $s_i=\downarrow$. Thus, \textsc{Ising}$\leq$\textsc{Hartree-Fock} under a Karp reduction. 

The promise on the error tolerance, $\delta$, given in the specification of the problem is necessary because of the pseudo-spin representation used in~\eqref{eq:ZZ}. Corrections to the energy due to the pseudo-spin representation can be computed using perturbation theory with the unperturbed Hamiltonian given by~\eqref{eq:penalty} and the converted Ising Hamiltonian, c.f.~\eqref{eq:ZZ}, as the perturbation.  The first order corrections in small parameter $C^{-1}$ are the Ising energies and the errors due to the pseudo-spin mapping arise at second order in $C^{-1}$.  A coarse estimate for $C$ is obtained by multiplying the number of non-zero terms by the maximum absolute value of a coefficient occurring in~\eqref{eq:2f}. Since we map the problem to \textsc{Ising}, there are $O(N^2)$ terms in the summation and the maximum absolute value of each term is unity. Hence, $C$ is estimated as $O(N^2)$. So long as $\delta<O(N^{-2})$, the first order corrections occurring at order $C^{-1}$ can be distinguished and the ground state Ising energies can be recovered.

To prove inclusion of HF in the complexity class NP, an algorithm for {verifying} the energy in polynomial time must be given.  If the coefficient matrix $C$ describing the correct orbital rotation is given, then the energy is calculated easily using Slater-Condon rules~\cite{Levine06}. Thus, given the answer, the accept/reject conditions are quickly verified.  Since the problem can also be quickly verified, \textsc{Hartree-Fock} is NP-complete.

\section{2-RDM methods}\label{sec:reduced}
Many computational chemistry algorithms seek to minimize the energy or other properties by manipulating the wave function but to quote \citet{Coulson60}, ``wave functions tell us more than we need to know\ldots All the necessary information required for energy and calculating properties of molecules is embodied in the first and second order density matrices.'' Extensive work has been done to transform this remark into a host of computational methods in quantum chemistry~\cite{Mazziotti07,Parr89}.  In this section and the following, we review the prominent computational complexity results related to this body of work.  

With respect to the 2-RDMs, it is easy to evaluate properties of the electronic system since the Hamiltonian only contains two-electron terms.  However, difficulties arise when determining if the 2-RDM is valid or invalid.  Unconstrained optimization of the energy using 2-RDM can lead to ground state energies of negative infinity if the validity of each 2-RDM cannot be determined.  While the criteria for validity have recently been developed~\cite{Mazziotti12a,Mazziotti12b}, deciding the validity of each 2-RDM using such criteria is known as the \emph{$N$-representability} problem. \citet{Mazziotti12a} also provides a proof that the following problem, \REP, is at least NP-hard.  In the next subsection, we follow~\citet{Liu07} to demonstrate the stronger conclusion that \REP~is QMA-complete. 

\paragraph*{Computational Problem: }~\REP.  The inputs are the number of electrons, $N$, an error tolerance $\beta\geq 1/\textrm{poly}(N)$, and a 2-RDM, $\mu^{(2)}$.  The task is to decide ($i$) if $\mu^{(2)}$ is consistent with some $N$-electron mixed state, $\rho^{(N)}$, or ($ii$) if $\mu^{(2)}$ is bounded away from all 2-RDMs that are consistent with an $N$-electron state by at least $\beta$~\footnote{The appropriate metric is the trace distance, $d_{tr}(A,B)=\|A-B\|_{tr}=\frac12 Tr[\sqrt{(A-B)^\dag (A-B)}]$, which generalizes the distance metric from standard probability theory.} promised that either $(i)$ or ($ii$) is true.

\subsection{\REP~is \qma-complete}
To show QMA-hardness, a Turing reduction is used to show \tLSH $\le$ \REP.  Before proceeding to the Turing reduction, note, since \REP~only considers a fixed number of electrons, the one-electron operators are not needed as $\ad_ia_j=\ad_i\left( \sum_k^M\ad_ka_k \right)a_j/(N-1)$. Each valid 2-RDM can be represented by an ${M\choose 2} \times{M\choose 2}$ dimensional matrix with ${M\choose 2}^2-1$ independent parameters\footnote{Since each 2-RDM is hermitian, the complex off-diagonal elements occur in pairs and the diagonal elements must be real.  Since the trace is normalized to unity, there a reduction of one degree of freedom, leaving ${M\choose 2}^2-1$ independent parameters.}.  A complete set of observables, such as the projection onto each matrix element\footnote{\citet{Liu07} used a different set observables inspired by the Pauli matrices with more convenient mathematical properties.}, is then used to characterize the space of 2-RDMs.

The space of valid 2-RDMs is convex; that is if $\mu^{(2)}_1,\mu_2^{(2)},\cdots, \mu_L^{(2)}$ are valid 2-RDMs, then the convex sum $\sum_jn_j \mu^{(2)}_j$ with $\sum_j^L n_j=1$ is also a valid 2-RDM.  This follows directly from the convexity properties of sets of density matrices and probability distributions. 

With access to an oracle for \REP, the boundaries of the convex set of valid 2-RDMs can be characterized. Because the oracle is used multiple times throughout the verification procedure, this is a Turing reduction. Since convex minimization problems can be solved efficiently and reliably, see e.g.~\cite{Boyd04}, the QMA-hardness is nearly demonstrated.  The last point of concern addressed by  \citet{Liu07}, are the errors introduced by the promise given on the oracle. To demonstrate that the algorithm remains robust in spite of such errors, the authors use a tailored version of the shallow-cut ellipsoid convex optimization technique~\cite{Grotschel88}.

Let us remark that the required error tolerance follows since the reduction relies on the QMA-completeness of \tLSH~which includes a promise on allowed the error tolerance.

To demonstrate that \REP~is QMA-complete, what remains is demonstrating that \REP~is in the complexity class QMA. If $\mu^{(2)}$ is a valid 2-RDM, then, relying on the Jordan-Wigner transform introduced earlier, Merlin can send polynomial copies of the correct $N$-electron state, $\mu^{(N)}$. The verifier, Arthur, first checks that the number of electrons in the given state is $N$ by evaluating $\textrm{Tr}[\mu^{(N)}\sum a_k^{\dagger}a_k]$.  Then Arthur randomly picks observables from the complete set described earlier and tests that the expectation value of the state $\mu^{(N)}$ and $\mu^{(2)}$ match until convinced. 
Merlin might try to cheat by sending entangled copies of $\mu^{(N)}$ but it can be proven that Arthur cannot be fooled based on the Markov inequality\footnote{Given any random variable $X$, the Markov inequality states $\langle |X|\rangle\geq a\textrm{Pr}(|X|\geq a)$ with $\langle X\rangle$ defining the expectation value of random variable $X$ and $|X|$ defining the absolute value of $X$.}.  See \citet{Aharonov03} for a proof of this fact.

\subsection{Restriction to pure states} 
The pure state restriction of \REP~has interesting consequences.  Consider,  \textsc{Pure} \REP, where the question is now: ``Did the 2-RDM come from a $N$-electron pure state (up to error $\beta$)?'' This contrasts with the original problem where the consistency questions refers to an $N$-electron mixed state. 

This problem is no longer in QMA since the verifier, Arthur, cannot easily check that the state is pure if the prover, Merlin, cheats by sending entangled copies of $\ket{\Psi^{(N)}}$.  If instead, two independent unentangled provers send the proof state, the state from the first prover can be used to verify the purity. The purity is checked using the pairwise swap test\cite{Harrow10}, a quantum algorithmic version of the Hong-Ou-Mandel effect in quantum optics\cite{Hong87}, on each of the supposedly unentangled states.  When the given state is not a product state (or nearly so), Arthur will detect it.  If the set of states from the first prover pass the test, the second set can be used to verify that randomly selected expectation values of $\ket{\Psi^{(N)}}$ and $\mu^{(2)}$ match.  The complexity class where two verifiers are used instead of one verifier is QMA(2).

\section{Density functional theory}

In this section, a further reduced description is examined: density functional theory. Density functional theory (DFT) methods are of profound importance in computational chemistry due to their speed and reasonable accuracy~\cite{Koch01,Parr89,Kohn99}. In this section, the complexity of the difficult aspects of DFT is shown to be QMA-complete. This was first conjectured by~\citet{Rassolov08} and rigorously proven by~\citet{Schuch09}. 

In DFT, the wave function is replaced by the one-electron probability density, $n(\x)=\rho^{(1)}(\x,\x)$.  The use of an object in three spatial dimensions to replace an object in $3N$ spatial dimensions without losing any information seems almost absurd, but the theoretical foundations of DFT are well established~\cite{Hohenberg64,Parr89}. The Hohenberg-Kohn theorem~\cite{Hohenberg64,Kohn99} proves that the probability density  obtained from the ground state wave function of electrons in a molecular system is in one-to-one correspondence with the external potential usually arising from the static nuclear charges.  Therefore, all properties of the system are determined by the one-electron probability density.

In the proof of the Hohenberg-Kohn theorem, one of the most important, and, elusive, functionals of the one-electron density is encountered: the universal functional.  In all electronic Hamiltonians, the electrons possess kinetic energy and the electrons interact via the Coulomb interaction; only $V_{eN}$ due to the nuclear configuration changes from system to system.  Separating these parts allows one to define the universal functional of DFT as
\begin{equation}
  F[n(\x)]=\min_{\rho^{(N)}\rightarrow n(\x)} \textrm{Tr}[(T_e+V_{ee})\rho^{(N)}],
  \label{eq:universal}
\end{equation}
which takes as input the probability density and returns the lowest possible energy of $T_{e}+V_{ee}$ consistent with the probability density.  The nuclear potential energy can be directly determined efficiently using $\int d\x \;n(\x) V_{eN}(\x)$ using Gaussian quadratures or Monte Carlo sampling.  Regardless of the method used to evaluate the integral, the domain does not (explicitly) increase with the number of electrons.  As a decision problem, we have the following:

\paragraph*{Computational Problem: }~\UF. The inputs are the number of electrons, $N$, a probability density $n(\x)$, a trial energy $E_T$ and an error tolerance $\delta< 1/\textrm{poly}(N)$. The task is to decide if $F[n(\x)]$ is greater than $E_T+\delta$ or less than $E_T-\delta$ promised that $F[n(\x)]$ is not between $E_T\pm\delta$. It is required that the summation over the Hamiltonian coefficients $\sum |V_{ijkl}^{ee}|+|T_{ij}^e|+|V^{eN}_{ij}|+|V^{mag}_{ij}|$ scale less than poly($N$).
	
\subsection{\UF~is \qma-complete}\label{sec:DFTQMA}
The demonstration that \UF~is QMA-hard proceeds via a series of reductions. Specifically, \citet{Schuch09} show that $H_{2QMA}\leq H_{heisenberg}\leq H_{hubbard}\leq H_{elec}$ where $H_1\leq H_2$ means that instances of the ground state decision problem for Hamiltonian $H_1$ can be embedded (Karp reduced) into ground state problem instances of Hamiltonian $H_2$. In their proof, the authors utilize the magnetic field to encode the problem instances causing $H_{elec}$ as defined in \eqref{eq:elec} to include an additional local magnetic field, $V^{mag}$. Note that the local field only affects the spin of the electron and does not require spin-dependent density functionals since the charge and spin do not couple. The Hamiltonian, $H_{2QMA}$, was listed earlier in \eqref{eq:kQMA}. Again since the reduction relies on the QMA-completeness of the \tLSH~problem, the promised error tolerance and the upper-bound of the coefficients of the Hamiltonian are required for QMA-completeness.

The $H_{hubbard}$ and $H_{heisenberg}$ Hamiltonians are commonly encountered models in condensed matter physics and are of the form
\begin{eqnarray}
	H_{hubbard}&=&\sum_{\langle i j \rangle}^{M/2}\sum_{s\in\{\uparrow,\downarrow\}} ta_{i,s}^\dagger a_{j,s}+\sum_i^{M/2}U \ad_{i\uparrow}\ad_{i\downarrow}a_{i\downarrow}a_{i\uparrow}\nonumber\\
	&&-\sum_i^{M/2}\sum_{d\in\{x,y,z\}}B^d_i\sigma_i^d\label{eq:hub}\\
	H_{heisenberg}&=&\sum_{d\in\{x,y,z\}}\sum_{ij}^{M} J\sigma^d_i\sigma^d_j +\sum_i^{M}B^d_i\sigma_i^d\label{eq:heis}
\end{eqnarray}
The first Hamiltonian describes an electronic system where the Pauli matrices, $\sigma^d$, are expressed using orbital pairs as in~\eqref{eq:fermionicZ}, while the second Hamiltonian describes a system of quantum spin. In both Hamiltonians, different problem instances are embedded using the local magnetic field. The embedding of $H_{heisenberg}$ instances into $H_{hubbard}$ follows directly from the orbital pair mapping described earlier in section~\ref{sec:relate}.  The remaining two Karp reductions are more involved.

First, let us consider the reduction $H_{2QMA}\leq H_{heisenberg}$.  This Karp reduction follows along the same lines used in reducing $kQMA$ from $k=5$ to $k=2$ based on \emph{perturbative gadgets}~\cite{Oliveira08,Kempe06} widely used in quantum complexity proofs. A mediator spin splits the state space of the system into a low energy and a high energy sectors. In the low energy sector, the perturbative coupling to the high energy states ``mediates'' new interactions in the low energy sector. For example, with $H_m=B_m\ket{\varphi_m}\bra{\varphi_m}$ acting on the mediator spin and a perturbation $V=\sigma^X_i\sigma^X_m\sigma^0_j+\sigma^0_i\sigma^Y_m\sigma^Y_j$, in the low energy space of spin $m$, there is, at second order, an effective interaction: $\sigma^X_i\sigma^Y_j$. Since the Hamiltonian problems refer to the ground state energy, the high energy sector is not important.

The reduction $H_{2QMA}\leq H_{heisenberg}$ requires:
\begin{enumerate}
	\item converting arbitrary strength couplings to constant strength couplings: $J_{12}\sigma^A_1\sigma^B_2\mapsto J(\sigma^A_1\sigma^M_m+\sigma^N_m\sigma^B_2)$;
	\item converting inequivalent couplings to equivalent couplings: $\sigma^A_1\sigma^B_2\mapsto\sigma^A_1\sigma^A_m+\sigma^B_m\sigma^B_2$;
	\item and finally converting equivalent couplings to Heisenberg interactions: $\sigma^A_1\sigma^A_2\mapsto \sum_d\sigma^d_1\sigma^d_m+\sum_d\sigma^d_m\sigma^d_1$.
\end{enumerate}
The full transformation requires 15 mediator spins where the local field on each of the mediator spins splits the system into low and high energy sectors. The specific local field depends on the interaction desired. The strength of the local field applied to the mediator spin $|B_m|$ is usually very large to ensure that perturbation theory applies.  So far this has limited the practical relevance of these constructions; for instance, field strengths required on the final set of mediator spins scales at nearly the 100th power of the system size.

The remaining reduction, $H_{hubbard}\leq H_{elec}$, was heuristically known since the Hubbard model phenomenologically describes electrons in solid state systems. \citet{Schuch09} rigorously demonstrate this reduction by careful accounting for error terms. They begin with a simple model used for studying solids where non-interacting (spin-less) electrons are subjected to a periodic delta function potential with $M/2$ such sites. The orbitals of this system can be solved exactly. The resulting second quantized Hamiltonian has uniform kinetic hopping terms: $\sum_{\langle ij \rangle}t a_i^\dag a_j$.  After incorporating the electron spin, electron-electron interactions are introduced. The strength of this interaction is rescaled by changing the spatial distance between neighboring sites until only the electrons at the same site can interact. Since each of the $M/2$ identical sites can only support one bound state, the exchange integral must vanish and the Coulomb integral is the same for each interaction yielding $\sum_{i}U\ad_{i\uparrow}\ad_{i\downarrow}a_{i\downarrow}a_{i\uparrow}$. 

Since the Hubbard model and the Heisenberg model require magnetic fields to define problem instances, the electronic Hamiltonian which they consider is not precisely the same as \eqref{eq:elec}; this also requires that the functional takes into account the separate spin components of the probability density when performing the minimization in \eqref{eq:universal}. 

Regardless, a polynomial time solution of \UF~would also solve the QMA-hard electronic Hamiltonian with local magnetic fields. This follows as the functional is convex
\begin{eqnarray}
	F[\sum p_jn_j]&=& \min_{\rho^{(N)}\rightarrow\sum p_jn_j} \textrm{Tr}\left[ (T_e +V_{ee})\rho^{(N)} \right]\nonumber\\
	&\leq& \sum_j p_j \min_{\rho_j^{(N)}\rightarrow n_j} \textrm{Tr}\left[ (T_e +V_{ee})\rho_j^{(N)} \right]
\end{eqnarray}
and the one-electron probability densities are also convex allowing one to perform convex optimization to find the minimum energy of the QMA-hard electronic Hamiltonian with the local magnetic fields.  Since the conditions for consistency are simple to check for one-electron probability densities\cite{Parr89}, the complications encountered in 2-RDM convex optimization from the approximate consistency conditions are not present. 

The inclusion of \UF~in complexity class QMA follows along the same lines of the generic two-fermion problem discussed in section~\ref{sec:generic}. However, in this case, Arthur must subtract the energy of the local field from the total energy to decide if the energy of $F$ is above or below $E_T\pm \delta$.

\subsection{Restriction to pure states}
The pure state restriction of DFT affects the complexity of computing $F$. Consider evaluating $F[n]$ where the input density arises from optimization over pure states $\ket{\Psi}$ instead of mixed states as in \eqref{eq:universal}. In this case, the universal functional is no longer convex. $$\sum_i p_i\min_{\Psi_i\rightarrow n_i}\bra{\Psi_i}(T_e+V_{ee})\ket{\Psi_i}\neq\min_{\Psi\rightarrow\sum p_i n_i}\bra{\Psi}(T_e+V_{ee})\ket{\Psi}.$$ The optimization problem would then be NP and an oracle for {\sc Pure} \UF~would allow a Turing reduction from QMA to NP.  Stated differently, an NP machine with access to an oracle for {\sc Pure} \UF, could solve any problem in QMA. Similar to the 2-RDM case,  the restriction to pure states complicates the verification such that two Merlins are required to show that the state is pure and that the proposed solution is correct.

\section{Other topics}
In this article, we have focused on three pillars of quantum chemistry but there are several other results worth mentioning.  \citet{Brown10complexity} showed that calculating the density of states for quantum systems is in the same complexity class as computing the classical partition function.  The complexity class of these problems is \#P where the problem instances request the number of solutions to problem in complexity class P. \citet{Rassolov08} briefly examined the complexity of the de Broglie-Bohm formulation of quantum mechanics finding that Monte Carlo techniques can be employed to efficiently sample the propagated Hamiltonian-Jacobi equations without the quantum potential.  They then suggest that the singularities of the quantum potential will ultimately be the source of difficulties in the quantum propagation. \citet{Troyer05} showed that the sign problem, properly defined, arising when evaluating quantum partition functions with Quantum Monte Carlo is NP-hard. Extension of DFT can be used in the time dependent domain in what is termed time-dependent density functional theory (TDDFT). \citet{Tempel12} recently demonstrated that one can construct TDDFT functionals for use in quantum computation. 

\section{Concluding remarks}

To close the paper, we make a few remarks on scaling of the error tolerances.  In the computational problems investigated, the error tolerance was upper bounded by an inverse polynomial $1/q(N)$ in the system size. In \HF, it was the pseudo-spin mapping that necessitated the promise, but, in \UF~and \REP, the promise is inherited from the \tLSH~problem. At first glance, it would appear that the error tolerance becomes more restrictive as the system size increases, however this is not necessarily the case.  

We focus on the problem \kLSH~since it is at the heart of the reductions for both \UF~and \REP.  In this problem, the error tolerance, $\delta$, is normalized by the number of terms in the Hamiltonian\footnote{One may ask why not use the norm of the Hamiltonian? For a Hamiltonian, $H$, the operator norm is the maximum eigenvalue and computing the maximum eigenvalue is just as hard problem as computing the ground state of $-H$, thus also a difficult problem.}, $m$. Therefore, the error tolerance only shrinks with system size if the number of non-zero terms in the Hamiltonian is constant or grows slower than $q(N)$. In many cases, the number of terms in the Hamiltonian increase with system size. Thus, even if asking for fixed error, e.g.~1 kcal/mol, as the system size increases the promise is still fulfilled and the problem of deciding instances of \kLSH~is QMA-complete.

When the error tolerance is independent of the system size or grows slowly, it is not clear if the problem remains QMA-complete.  This is related to the on-going research into possible quantum generalizations of the PCP theorem\footnote{The term PCP refers to probabilistically checkable proof systems where the verifier is only allowed to randomly access parts of a non-deterministically generated proof.  Using only polynomial time and a fixed amount of random bits, the verifier is to validate or invalidate the proof.}    The PCP theorem shows that the task of approximating some NP-complete problems is also NP-complete~\cite{Arora1998a,Arora1998b}. A quantum generalization, if it exists, would allow one to show that \kLSH~remains QMA-complete even when the normalized error tolerance is bounded by a constant instead of an inverse polynomial of $N$. This remains a prominent open question in modern computer science and is discussed more thoroughly by~\citet{Osborne12}.

With respect to complexity in quantum chemistry, we have examined three computational complexity of different quantum chemistry algorithms: \HF, \REP, and \UF.   The difficult of these computational chemistry problems relative to problems found in other fields gives insights into why computational chemistry is difficult.  The difficulty of these problems imply that even a quantum computer will be unable to solve all instances since it is believed that BQP does not contain NP. However, even if quantum computers are never constructed, the study of quantum computational complexity gives new insight into the relative difficulty of problems.  For instance, characterizing valid two-electron reduced density matrices or evaluating the universal functional of DFT is arguably more difficult than evaluating the Hartree-Fock energy based on the probable separation of complexity classes QMA and NP.  The examination of other properties of molecular systems and other problems encountered in quantum chemistry will provide fertile ground for future research into the difficulty of quantum chemistry. 

\section*{Acknowledgements}
JDW and PJL acknowledge support from the NSF under award numbers~1017244 and PHY-0955518, respectively. PJL and AAG acknowledge support from the NSF CCI, ``Quantum Information for Quantum Chemistry,'' (CHE-1037992). AAG additionally thanks HRL Laboratories, The Dreyfus Foundation and The Sloan Foundation for their support. JDW would like to thank The Max-Planck Institute PKS, Dresden and ISI, Torino where parts of this work were completed. The authors wish to thank Z. Zamboris, S. Aaronson, J. Biamonte, D. Gavinsky, and M.-H. Yung for helpful comments on the manuscript. JDW also acknowledges helpful discussions with T.~Ito about the PCP theorem.

\footnotesize{
\providecommand*{\mcitethebibliography}{\thebibliography}
\csname @ifundefined\endcsname{endmcitethebibliography}
{\let\endmcitethebibliography\endthebibliography}{}

}

\end{document}